\documentclass[12pt]{iopart}
\usepackage{graphicx}

\newcommand{\gi}{\textsl{\textrm g}}
\newcommand{\muG}{$\mu$Gal}
\newcommand{\com}{COMSOL\textsuperscript{\textregistered}}

\begin{document}
\title{Self-attraction effect and correction on three absolute gravimeters}
\author{E. Biolcati$^1$, S. Svitlov$^2$, A. Germak$^1$}
\address{$^1$Istituto Nazionale di Ricerca Metrologica INRIM, Turin, Italy}
\address{$^2$Institute of Optics, Information and Photonics, University of Erlangen-Nuremberg, Erlangen, Germany}
\ead{e.biolcati@inrim.it}

\begin{abstract}
The perturbations of the gravitational field due to the mass distribution of an absolute gravimeter have been studied. The so called Self Attraction Effect (SAE) is crucial for the measurement accuracy, especially for the International Comparisons, and for the uncertainty budget evaluation. Three instruments have been analysed: MPG-2, FG5-238 and IMGC-02. The SAE has been calculated using a numerical method based on Finite Element Method simulation. The modelled effect has been treated as an additional vertical gravity gradient. The Self Attraction Correction (SAC) to be applied to the computed \gi\ value is of the order of $1 \times 10^{-8}$~m/s$^2$. 
\end{abstract}

\vspace{2pc}
\maketitle

\section{Introduction}
Modern transportable Absolute Gravimeters (AGs) measure the local value of the free-fall acceleration \gi\ using the reconstructed trajectory of a falling object in vacuum. The mass of the parts constituting each AG apparatus (such as laser interferometer, vibration isolation system, vacuum chamber, etc.) are sources of an additional gravitational field, which can systematically perturb the motion of the object. The so called Self Attraction Effect (SAE), as demonstrated in~\cite{robert,ago1}, is greater than 1 \muG\ (1 \muG\ = $10^{-8}$ ms$^{-2}$), which is not negligible in the uncertainty budgets of modern AGs.

The knowledge of the SAE and the specific self-attraction correction (SAC) is crucial for each measurement carried out by the AGs, because it can improve the measurement accuracy. It becomes very important when the measurements are used to calculate the Key Comparison Reference Value (not physically known) during the dedicated International Comparisons.

To calculate the SAE, a detailed study on three different gravimeters MPG-2~\cite{mpg2}, FG5-238~\cite{fg5} and IMGC-02~\cite{imgc2} has been performed. A Finite Element Method (FEM) simulation to calculate the contributions of each part of the gravimeters, as proposed in~\cite{ago1,dq}, has been used. To simplify calculation of the SAC, the SAE has been treated as an additional constant gravity gradient. 

The paper is structured as follows. In section~\ref{sec:2} a brief description of the three instruments is presented. Section~\ref{sec:3} describes the adopted method used to evaluate the SAE and the results obtained for the AGs. The SAC values are calculated separately for the three AGs in section~\ref{sec:4}. The uncertainty of the SAC is discussed in section~\ref{sec:5}, whilst the main conclusions are drawn in section~\ref{sec:6} .  

\section{Absolute gravimeters}\label{sec:2}
Three different transportable AGs have been studied:
\begin{itemize}
\item the MPG-2, designed in Germany by the Max Planck Institute for the Science of Light (MPL), prototype instrument;
\item the FG5-238, a commercial instrument produced by the U.S.A. Micro-g LaCoste Incorporation (the results coming from this AG can be considered for to the other instruments of the same type, i.e. FG5-2$xx$ with vertical legs of the supporting tripod);
\item the IMGC-02, developed in Italy by the Istituto Nazionale di Ricerca Metrologica (INRIM), prototype instrument.
\end{itemize}

For each of them, the measurement of the \gi~value is obtained using the reconstructed trajectory of a corner-cube prism (or a hollow retroreflector) which moves vertically in vacuum. The IMGC-02 takes into account for both the rise and fall motions of the flying object, whilst the other two instruments measure the acceleration during the free-fall motion only.   

Automated systems are employed to centre, launch and receive the object event by event with nominal rates of about (0.02~-~0.1)~Hz during data taking sessions of several hours. An interferometric system is implemented to obtain time and distance coordinates of the trajectory using a visible laser beam. The interferometer measures the distance between the free falling corner cube test mass and a second retroreflector mounted on the quasi-inertial mass of a vibration isolation system (which is used to isolate it from ground vibrations). A detailed description of the three AGs can be found in~\cite{mpg2,fg5,imgc2}. 

The distribution of the heaviest and nearest parts of each instrument and the path of the flying object have been considered to characterise the perturbing gravitational field $\Gamma(Z)$. Each gravimeter can be essentially divided in three main parts:
\begin{itemize}
\item \emph{read-out electronic case:} it can be easily moved farther than 1~m from the flying object trajectory, so the SAE from this source is negligible\footnote{As an example, a mass of 200~kg, at least twice bigger than typical AG electronics, placed at 1~m from the trajectory axis produces an effect along the vertical direction less than 0.1~\muG.};
\item \emph{measuring system:} supporting tripod, seismometer or super-spring system with its own support, several detectors and the interferometer;
\item \emph{launch system:} the vacuum chamber with the dropping mechanism and its basis with all the accessories, in this case the effect must be analysed in details because those objects are very close to the flying object trajectory. 
\end{itemize}

\begin{figure}[!htbp]
\begin{indented}\item[]
\includegraphics[width=0.67\textwidth,trim=0 20 0 10,clip]{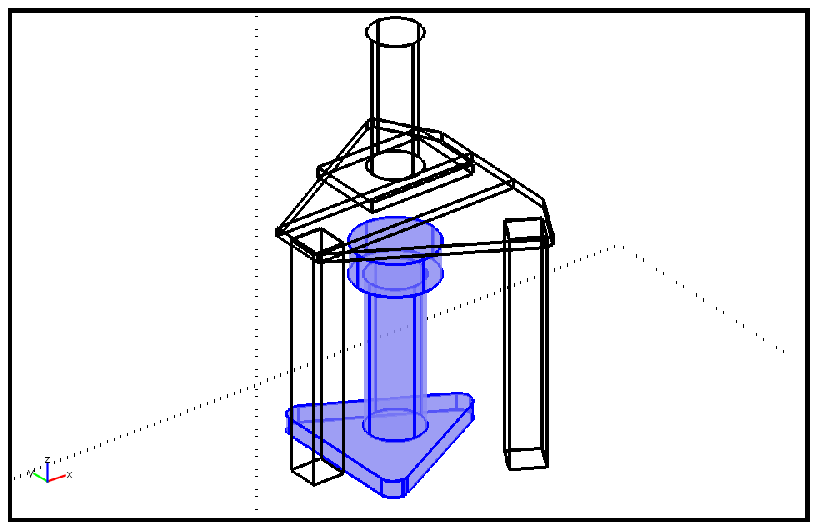}
\includegraphics[width=0.67\textwidth,trim=0 20 0 10,clip]{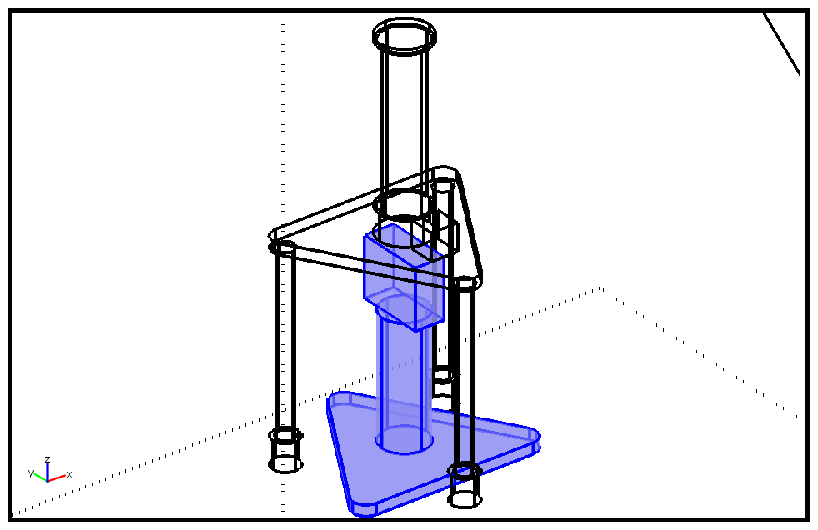}
\includegraphics[width=0.67\textwidth,trim=0 20 0 10,clip]{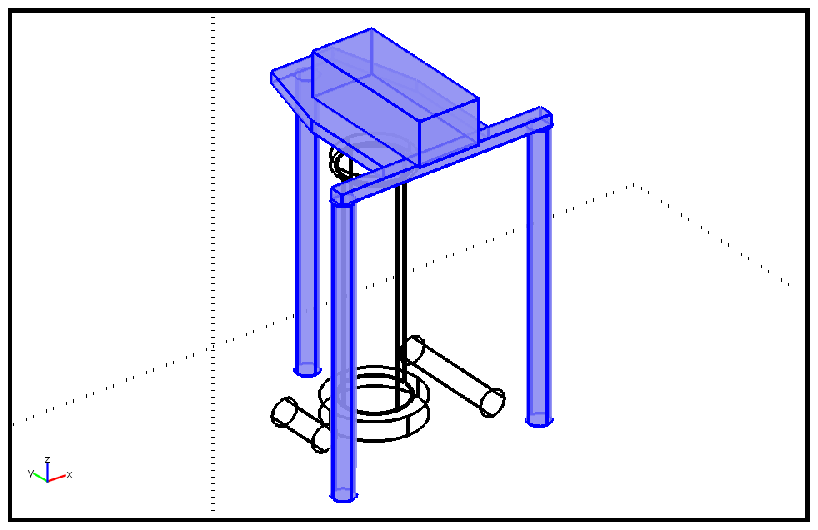}
\caption{Geometry of the MPG-2 (top), FG5-238 (middle) and IMGC-02 (bottom) absolute gravimeters, drawn using the \com\ 3D module. The two main systems are distinguishable: measuring (coloured) and launch one (transparent). Only parts important for the gravitational effect analysis have been simulated.}
\label{fig:2}
\end{indented}
\end{figure}

The geometry of the last two parts have been drawn using the \com\ 3D module and it is shown for the three AGs in figure~\ref{fig:2}. Screws, cables, small holes are not simulated because their influence is negligible for the SAE. 

In the free-fall AGs, a co-falling system is implemented. It consists of a support which moves ahead of the test mass to catch it at the end of the free fall~\cite{mpg2,fg5}. The distance between such system and the test mass varies during the trajectory from zero to few millimeters. For this reason a non constant SAE is present. It can not be included in the time-independent FEM simulation, hence an approximation of the average effect has been evaluated with the appropriate uncertainty.

\section{Self-Attraction Effect}\label{sec:3}
To evaluate the gravitational field perturbation of each single part of an AG, an accurate knowledge of the geometry and the mass distribution of the source is needed. Due to the complexity of the single parts (edges, different materials, non regular shape), a FEM simulation has been preferred instead of a mathematical modelling. 

The \com\ software~\cite{comsol} has been adapted to this purpose. A package dedicated to the gravitational effects is not implemented in the original software. The electrostatic module is then used, exploiting the analogy between gravitational and electrical interactions,  as already used for this or similar purpose in~\cite{ago1,dq}. 

The component along the direction orthogonal to the floor on which a gravimeter is located (called Z coordinate) of the gravitational field ($\Gamma_Z$) has been considered to evaluate possible effects on the measurement of the \gi\ value. The other two components along the X and Y axes can influence the flying object trajectory introducing rotation or shift components. This effort has not been treated in this study because the perturbations are expected to be negligible with a respect to the one along the Z~coordinate.  

The origin of the reference frame has been set on the ground, with the Z~axis directed upwards and crossing the centre of the flying object. Modern AGs measure the \gi\ value at the height values between about 0.4 and 1.3~m. The $\Gamma_Z$ value is calculated along a straight line from $(0,0,0)$ to $(0,0,1.5~{\rm m})$. A range larger than the distance covered by the free-falling object has been chosen in order to show the perturbation of $\Gamma_Z$ due to all the simulated parts.  

The FEM simulation consists of three main steps: geometry draw, mesh implementation and equations solution. 

In the first step, the parts of the AG are drawn and their mass values implemented. Two different approaches can be used for each part:
\begin{itemize}
\item if its shape is regular (such as cylinder or parallelepiped) and the mass distribution is almost homogeneous, the mass density and volume values are introduced in the software;
\item in case of complex shapes, e.g. different parts with screws and holes, or inhomogeneous mass distribution, as a seismometer, it is approximated to a regular solid with equivalent volume and mass.
\end{itemize}
A detailed study has been performed on a single object in order to prove that the two approaches are equivalent with a respect the gravitational field uncertainty required for this report. The result of this test has been omitted and a similar procedure can be found in~\cite{ago1}. 

For this study seismometers, frames, launch chambers, vacuum pumps have been simulated using the second approach. The other components have been simulated with the proper geometry and density parameters. 

The parts so defined are then embedded in empty volume where the field can be propagated. The shape is spherical with the radius value about one order of magnitude larger than the sizes of the AGs, in order to have edges far from the centre of the instrument, so as to minimise the boundary effects~\cite{ago1}. 

In the second step, the mesh geometry and size are implemented. Such parameters must be tuned to have the maximum resolution of the gravitational field in the range of interest. 

\begin{figure}[!htb]
\begin{indented}\item[]
\includegraphics[width=0.8\textwidth,trim=0 0 5 80,clip]{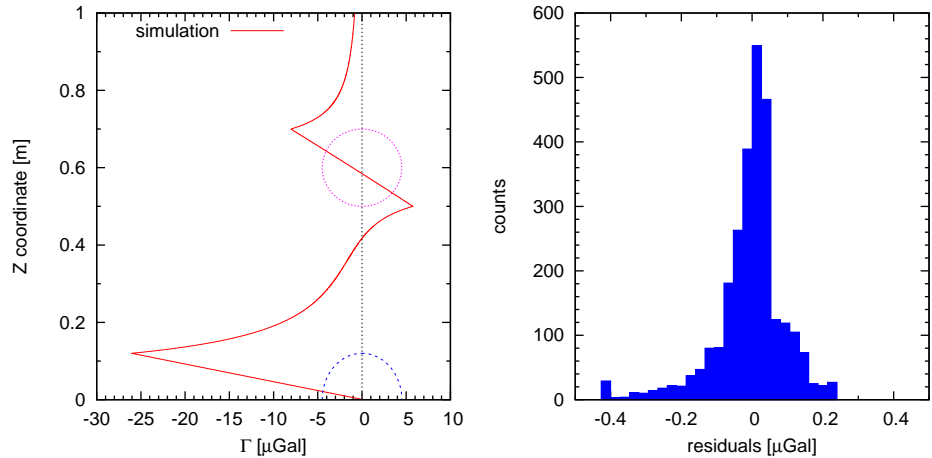}
\caption{Comparison between gravitational field values (component on Z coordinate, X=Y=0) obtained by using the theoretical formulas and the FEM simulation. Left: simulated field versus Z, steel and aluminum spheres also sketched. Right: residual distribution.}
\label{fig:1}
\end{indented}
\end{figure}

To validate the FEM simulation parameters, i.e. the mesh size and the boundary sphere, a study of two simple objects has been preliminary performed. A steel sphere centred at $(0,0,0)$ with radius $r_1=0.12$~m has been located together with an aluminium sphere centred at $(0,0,0.6~{\rm m})$ and $r_2=0.1$~m (figure~\ref{fig:1}). Materials, positions and sizes of the spheres have been chosen to cover the actual ranges of the SAE of several microgals. The radius of the boundary sphere is $r_0=3$~m. In figure~\ref{fig:1}~(left) the gravitational field (only component along Z coordinate at X=0 and Y=0) obtained from the FEM simulation is plotted. In the same figure the distribution of the residuals between values coming from theoretical model and simulation is plotted. A root mean square value of the residuals is 0.08~\muG\ with the mean about zero, figure~\ref{fig:1} (right). The 89.9\% of the residuals are below 0.1~\muG\ (absolute value). Such result will be considered as a measure of the discrepancy between the theoretical model and FEM simulation. 

In the last step of the method, the equations of the problem are numerically solved. In this case the only parameters can be set are referred to the CPU consumption for the algorithm, with modifications of the results below the 0.01\%~\cite{ago1}. 

The main contribution to the uncertainty of the simulation comes from the approximations made in the first step. In order to estimate it, several simulations of different complex parts (e.g. seismometer, launch chamber) have been performed varying the ratio between the sub-part masses\footnote{In other words, an object of known mass is constituted by sub-parts of unknown mass. Simulations with a ratio between the sub-parts set to 1/3, 1/3, 1/3 is performed. Then it is repeated with different ratio values as 1/8, 3/8, 1/2 etc.}. Using such approach, an uncertainty of 0.1~\muG\ has been estimated. 

The method has been applied to the three AGs with equal FEM parameters. A mesh called \emph{finer} (\com\ notation) has been implemented. The mesh has been adapted to the edges and the contact points of the objects to have about 300~k finite elements for each AG. The gravitational field has been calculated in a boundary sphere. The component Z of the gravitational field along the line centred with the flying object trajectory has been extracted for the global simulation. The contributions due to the measuring and the launch systems have been separately estimated.

\begin{figure}[!htbp]
\begin{indented}\item[]
\includegraphics[width=0.62\textwidth]{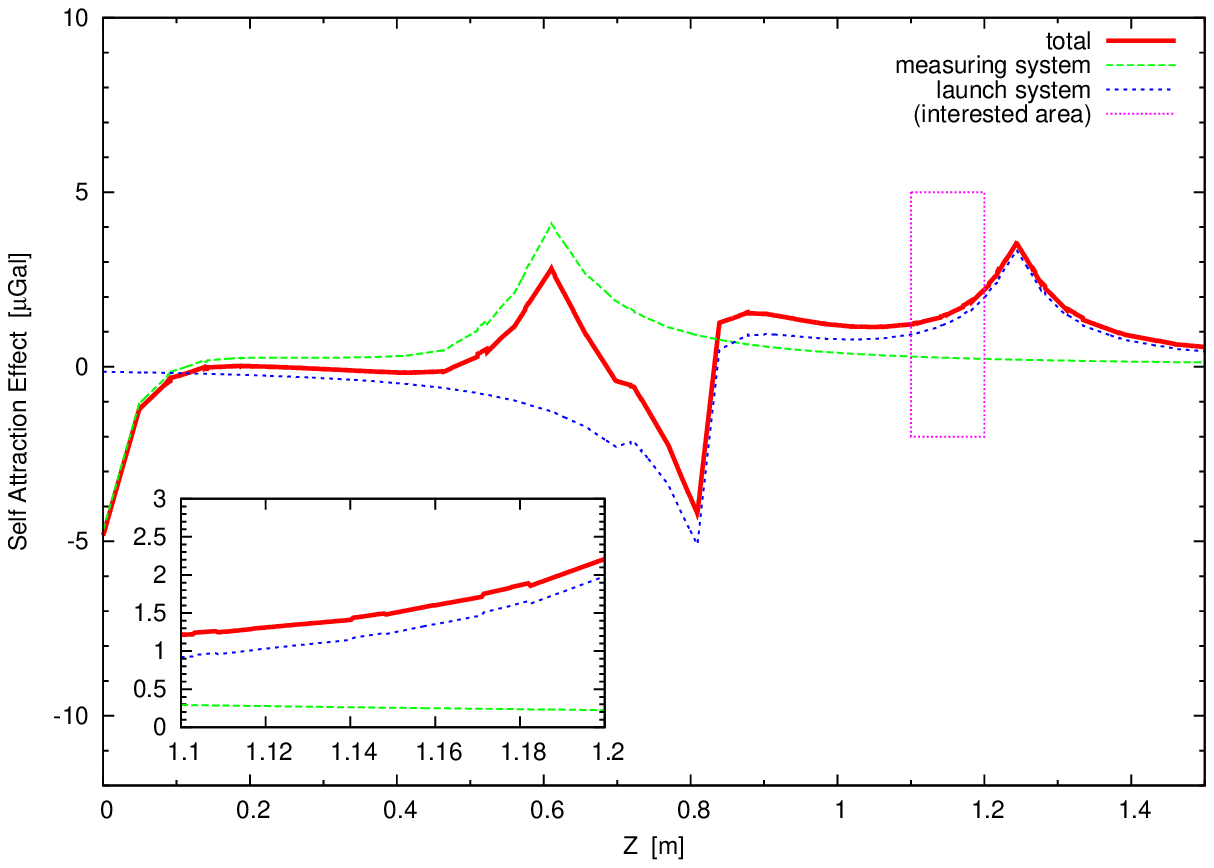}
\includegraphics[width=0.62\textwidth]{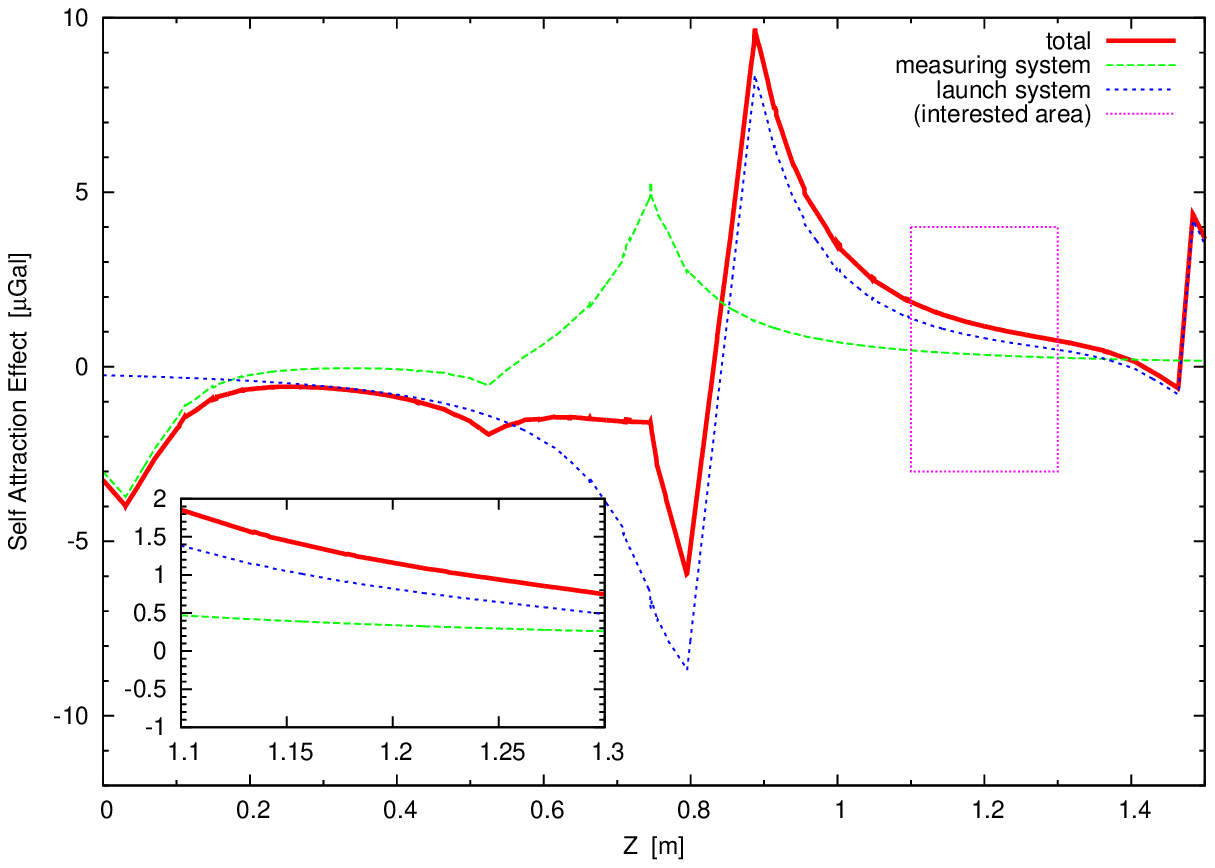}
\includegraphics[width=0.62\textwidth]{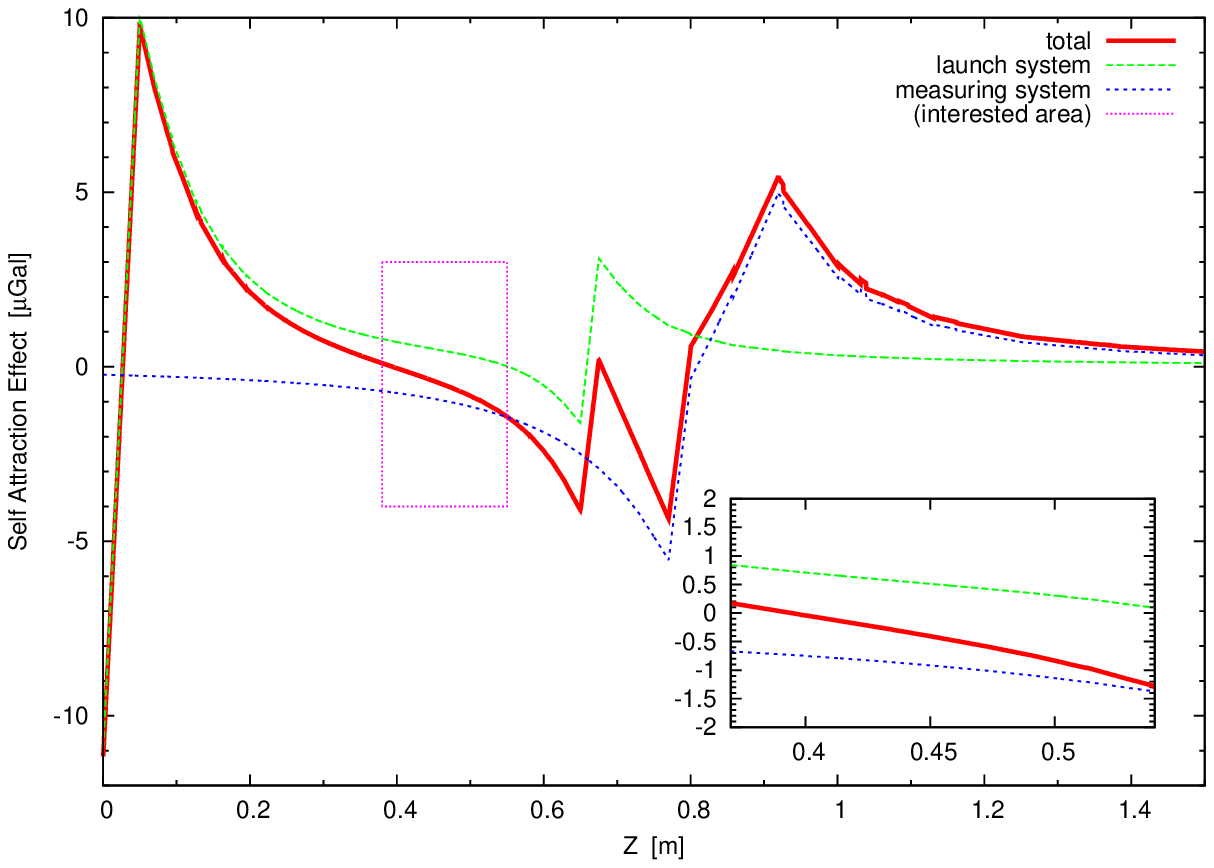}
\caption{Result of the FEM simulation for the absolute gravimeter MPG-2 (top), FG5-238 (middle) and IMGC-02 (bottom). Self-Attraction Effect versus Z coordinate. Total and single contributes of launch and measuring systems are shown. The approximate operating range of every instrument is identified by the rectangular in the figure and it is enlarged in the inset plot.} 
\label{fig:3}
\end{indented}
\end{figure}

In figure~\ref{fig:3} the total effect and contributions due to the individual instrument components are shown for the three AGs.

For the MPG-2, the largest values are located at $Z\simeq0.8$~m and $Z\simeq1.2$~m, due to the presence of the launch chamber basis. In the range of the free-falling path, a positive SAE below 2~\muG\ is observed and the behaviour appears monotone. Also for the FG5-238 the main effect can be observed at $Z\simeq0.8$~m, but in this case the operation range is larger than in the previous instrument ($\simeq0.2$~m). The SAE in such range is also below 2~\muG\ and positive. For the IMGC-02, two bumps can be observed at $Z\simeq0$ and $Z\simeq0.9$, due to the presence of the launch chamber basis and the seismometer with its support respectively. In the range of rise and fall path of 0.2~m, the SAE is in a range from --0.8~\muG\ to 0.4~\muG.

For the MPG-2 and FG5-238, an average contribution of ($0.1\pm0.1$)~\muG\ due to the co-falling system has to be added. Such value has been evaluated taking into account geometry and density of the support used to catch the test mass at the end of the trajectory and an average separation between the support and the test mass.

\section{Self-Attraction Correction}\label{sec:4}
\begin{figure}[!htbp]
\begin{indented}\item[]
\includegraphics[width=0.8\textwidth,trim=0 170 10 0,clip]{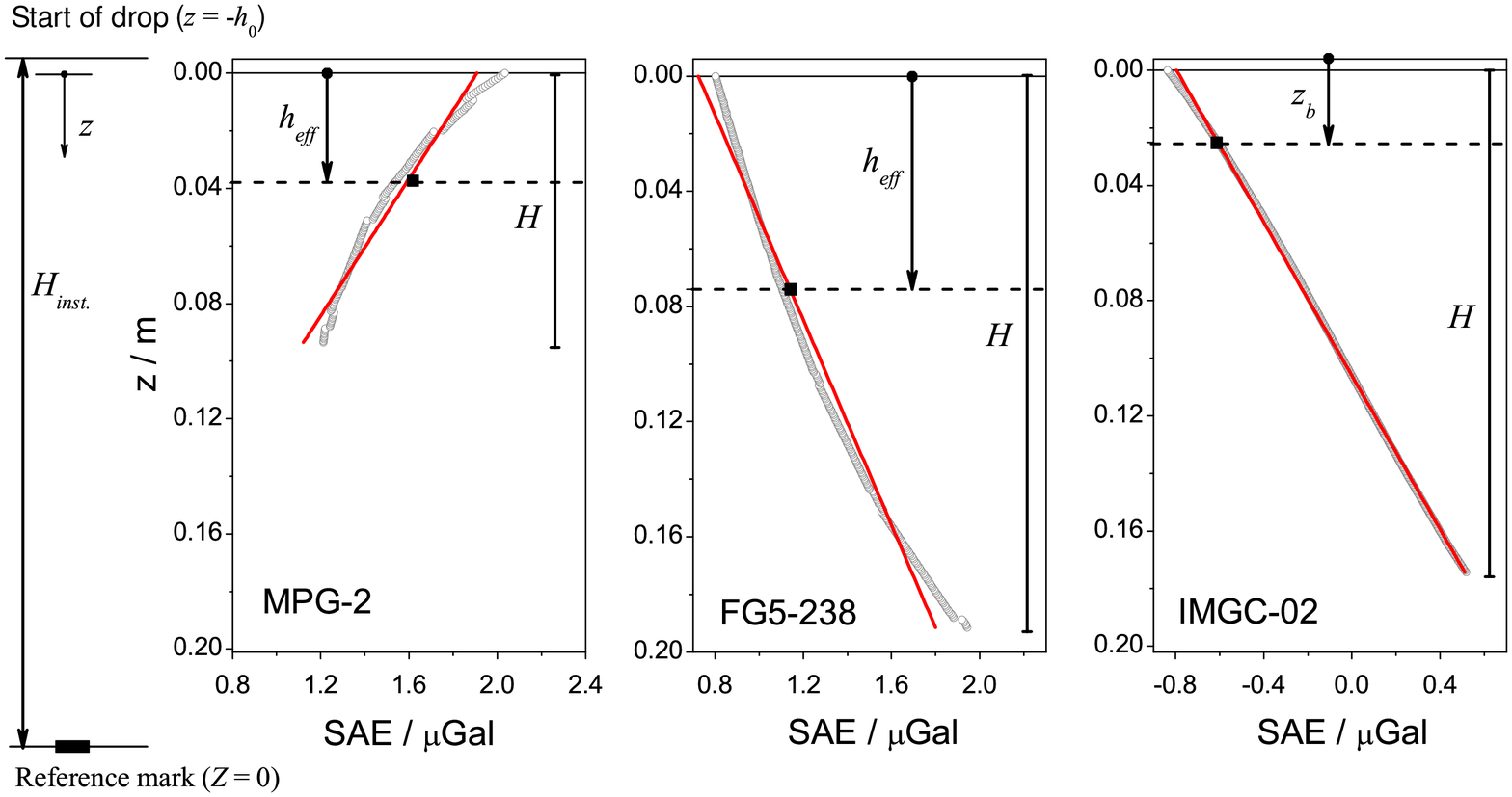}
\caption{Self-attraction effect over a measurement range $H$ of the MPG-2, FG5-238 and IMGC-02. Open grey dots for the SAE computed in previous section; red lines for the approximations by the first order polynomial. Solid squares indicate a value of the SAC (opposite sign) at the corresponding reference height. Computed corrections are shown in table~\ref{tab:serg}.}
\label{fig:serg}
\end{indented}
\end{figure}

The computed above SAE is approximately linear in the measurement range of every AG (figure~\ref{fig:serg}, when the straight line is fitted to the SAE, the coefficient of determination exceeds~0.9 for every instrument). Such linear gravity variation corresponds to the constant self-attraction gravity gradient~$\gamma_{\rm SAE}$ with a some constant offset~$g_{0\, \rm SAE}$ at a chosen origin, where the free-fall acceleration from the Earth gravity field equals $g_0$. The superposition property of gravitational fields from several sources allows to treat the SAE in an AG similar and in addition to the conventional vertical gradient $\gamma$ of the Earth gravity field. Then the perturbed height-varied free-fall acceleration is approximately given by 
\begin{equation}\label{eq:3}
g(z)=(g_0 +\gamma z)+(g_{0\, \rm SAE}+\gamma_{\rm SAE}z).
\end{equation}

The z-axis in~(\ref{eq:3}) is directed toward the centre of Earth (opposite to the Z~axis in the previous figures), and the signs of $g_0$ and $\gamma$ are chosen to be positive downwards (a freely-falling object accelerates downwards, magnitude of the acceleration increases towards the Earth surface). For the purpose of the SAC calculation, the z-origin is adjusted to the first measurement position (start of the total measurement distance interval~$H$, figure~\ref{fig:serg}).

Typically, AGs employ the linear free-fall motion model for the unperturbed free-fall~\cite{mpg2,fg5,torge}. The measurement result $g_m$ is obtained as the mean-weighted value of the perturbed free-fall acceleration (the weighting functions of different gravimeters are given in~\cite{nago1}). It equals to an instantaneous value of the linearly varied free-fall acceleration at the specific level below the level $z=0$ (start of $H$), called the effective measurement height $h_{\rm eff}$. Such parameter is independent on the constant gravity gradient and its uncertainty. It can be found analytically for different gravimeters, as shown in~\cite{nago1,nieb2,timmen,nago2,12a}.

Then at $z=h_{\rm eff}$ we have from~(\ref{eq:3})
\begin{equation}\label{eq:4}
g_m=g(h_{\rm eff})=(g_0+\gamma h_{\rm eff})+(g_{0\,\rm SAE}+\gamma_{\rm SAE} h_{\rm eff}).
\end{equation}

The second brackets in~(\ref{eq:4}) is the additive systematic error, which appears in the measurement result due to the SAE. When taken with the opposite sign, it is the self-attraction correction:

\begin{equation}\label{eq:5}
\Delta g_{\rm SAC} = -(g_{0\,\rm SAE}+\gamma_{\rm SAE}\,h_{\rm eff}).
\end{equation} 

After the measurement result $g_m$ is corrected by~(\ref{eq:5}), we assume that the instrument is removed from the observation site. Then this corrected result can be transferred to the desired height level, using the conventional vertical gradient~$\gamma$.

Figure~\ref{fig:serg} illustrates the SAC calculation. The total measurement interval $H$ offsets from the start of drop by some distance interval $h_0$, which is different for different gravimeters (in the case of the IMGC-02, the start of drop corresponds to the apex of the trajectory parabola).

The MPG-2 is a multiple-level free-fall AG with the linear free-fall motion model~\cite{mpg2}. A SAC of --1.59~\muG\ is obtained by the linear interpolation of the SAE to the level $z= h_{\rm eff}$ (figure~\ref{fig:serg}, left). A slight deviation of the SAE from the straight line is considered as the systematic effect in the uncertainty calculation (see after). 

The FG5-238 is a multiple-level free-fall AG with the linear free-fall motion model, which includes the gravity gradient $\gamma$ as an external known quantity~\cite{fg5}. Such transformation of the linear model automatically shifts the mean-weighted measurement result~(\ref{eq:4}) from the level $z=h_{\rm eff}$ to the start of measurement interval $H$ with a correction of $\Delta g_{\gamma} = - \gamma \, h_{\rm eff}$. Since the $h_{\rm eff}$ does not depend on the constant gravity gradient and on its distortion by the SAE, the same correction~(\ref{eq:5}) is valid for both results reported at $z=h_{\rm eff}$ or $z=0$ (including the known, but uncertain gravity gradient $\gamma$ into the fitted model causes an additional associated measurement uncertainty of $u_{\gamma}\,h_{\rm eff}$,  which is irrelevant to the SAC and is not considered in this paper). The computed SAC is of --1.13~\muG\ (\ref{fig:serg}, centre). 

Such two SAC values have to be increased by 0.1~\muG\ due to the co-falling system SAE, i.e. it becomes --1.69~\muG\ for the MPG-2 and --1.23~\muG\ for the FG5-238.

The IMGC-02 is the multiple-level rise-and-fall AG with the non-linear model, which contains the gravity gradient $\gamma$ as an unknown parameter~\cite{bich}:
\begin{equation}\label{eq:7}
z(t)=z_0+g_a\left [\frac{(t-t_a)^2}{2}-\varphi\frac{(t-t_a)^3}{6}+\gamma\,\frac{(t-t_a)^4}{24}\right ],
\end{equation}
where $g_a$ is the free-fall acceleration, $t_a$ is the time at the apex of the trajectory parabola and $\varphi$ the friction coefficient of the residual air in the launch chamber.

Since this model is non-linear in the parameters $g_a$ and $\gamma$, the superposition principle for the parameter estimates is not valid, and \emph{a posteriori} correction for a time (or height) varying perturbation is not applicable in a strict sense. Instead, a straightforward approach to modify the registered time-distance coordinates before the non-linear least-squares adjustment might be used by taking into account \emph{a priori} computed SAE. However, obtained results might not be extendible to other measurement conditions or changed parameters of an instrument. Then such a modification is necessary at every new measurement session. 

In the IMGC-02 a measurement result is usually reported at the best measurement height $z_b$ below the apex of the parabola~\cite{bich}:
\begin{equation}\label{eq:8}
g_b=g_a+\gamma\, z_b,
\end{equation}

where $g_a$ and $\gamma$ are directly estimated from the trajectory data using the model~(\ref{eq:7}), and $z_b$ is computed from the elements of the variance-covariance matrix and also depends on a quantity of the upper portion of the trajectory removed from the least-squares fit. With typical parameters of the IMGC-02 the best measurement height is close to the effective measurement height $h_{\rm eff}= H/6$ derived for the rise-and-fall AG with a linear free-fall motion model and without removing the upper portion of the trajectory~\cite{nago1}:
\begin{equation}\label{eq:9}
z_b \simeq \frac{H}{5.8}
\end{equation}
with $H$ being a total distance interval traveled by a falling object twice.

To evaluate an impact of the SAE on the measurement result~(figure~\ref{fig:serg}, right), we assume that at the level $z_b$ the estimate $g_b$~(\ref{eq:8}) is not correlated with the estimate $\gamma$~\cite{bich}. This indicates analogy between $z_b$ and $h_{\rm eff}$. Then the bias of $g_b$ can be evaluated similar to that, which contributes to the measurement result at the level $z=h_{\rm eff}$~(\ref{eq:5}). Consequently, the SAC is given by

\begin{equation}\label{eq:10}
\Delta g_{\rm SAC}=-[g_{0\, \rm SAE}+\gamma\, (z_b-h_0)].
\end{equation}

The SAC is then computed as +0.61~\muG~(figure~\ref{fig:serg}, right).

\section{Uncertainty of the SAC}\label{sec:5}
If the errors of the computed SAE were independent random variables, the uncertainty of the SAC, as given by (\ref{eq:3}), could be estimated using the conventional uncertainty propagation procedures. Since in our case the errors are not entirely independent (this was noticed from separate investigations of the residuals in figure~\ref{fig:1}) and in addition the SAE deviates from the straight line (figure~\ref{fig:serg}), we avoid the analytical uncertainty propagation and consider the uncertainty components of the SAC in the following way.  

\begin{description}
\item{\it Not complete modelling.} The not complete knowledge of density and geometry of the parts of gravimeters induces some systematic error of the computed SAC. According to simulations shown in section~\ref{sec:3}, we assign 0.1~\muG\ for every instrument. 

\item{\it FEM simulation.} The discrepancy between the FEM simulation and the mathematical model depends on the magnitude and curvature of the SAE in a complicated way (not shown in this report). As the over-estimate, we assign the standard deviation of the single value of the SAE (0.1~\muG) to the uncertainty of the SAC, obtained by the linear interpolation (\ref{eq:5}), for all the three instruments. 

\item{\it SAE non-linearity.} We consider the maximal deviations of the SAE from the straight lines (figure~\ref{fig:serg}) and assign it (with the over-estimation) to the uncertainties of the computed SAC. This gives 0.13~\muG, 0.15~\muG\ and 0.06~\muG\ for the MPG-2, FG5-238 and IMGC-02, respectively. 

\item{\it Co-falling system.} We have evaluated the correction due to such peculiarity of the free-fall systems as --0.1~\muG\ with the associated uncertainty of $\pm$~0.1~\muG.
\end{description}

Combining together, we find the SAC uncertainty of 0.2~\muG, 0.2~\muG\ and 0.1~\muG\ for the MPG-2, FG5-238 and IMGC-02, respectively. Table~\ref{tab:serg} resumes the computed SAC values and uncertainties for the three AGs together with proper parameters as initial velocity, measurement range, effective height, initial offset.

\Table{\label{tab:serg} The self-attraction corrections for three absolute gravimeters. Initial offset $h_0$ is the distance from the start of drop to the start of the measurement interval $H$. The effective measurement height $h_{\rm eff}$ is counted from the start of $H$, and the best measurement height $z_b$ is counted from the apex of the parabola. The SAC is computed by the linear interpolation of the SAE to the level $h_{\rm eff}$ (or $z_b$).} 
\br
 & MPG-2 & FG5-238 & IMGC-02 \\
\mr
reference height above floor $H_{\rm inst}$ [m]& 1.154          &  1.292            &  0.475 \\
initial offset $h_0$ [m]                       & --0.005        &  --0.006          & --0.004 \\
measurement range $H$ [m]                      & 0.094          &  0.192            & 0.174$\times$2 \\
effective height $h_{\rm eff}$ or $z_b$ [m]    & 0.038          &  0.074            & 0.029 \\
start velocity  [m/s]                          & 0.31           &  0.35             & 1.84 \\
SAC [\muG]                                     & --1.7          &  --1.2            & +0.6 \\
$u_{\rm SAC}$ [\muG]                           & 0.2            &  0.2              & 0.1 \\
\br
\endTable

\section{Conclusions}\label{sec:6}
Three Absolute Gravimeters have been studied in order to calculate the Self-Attraction Effect due to the masses of their single parts, using a FEM simulation method. The correction to be applied to the final measure of \gi\ has then been calculated for each AG. It should be pointed out that besides magnitude and curvature of the observed SAE, the SAC numerically depends on the method of a free-fall, adopted motion model, number and method of data location and portion of the reconstructed free-fall motion trajectory used in the least-squares adjustment (all of this is accumulated in the specific weighting function~\cite{nago1} of an AG).

The MPG-2 presents the largest SAE. A SAC of ($-1.7\pm0.2$)~\muG\ can be applied to the reported \gi\ value. For the FG5-238 a similar SAE has been found. In this case a SAC of ($-1.2\pm0.2$)~\muG\ can be applied to the measurement result reported at the start of drop or at the effective measurement height. This correction is consistent with the results reported in~\cite{robert}. For the IMGC-02 all range of the SAE appears to be negligible with respect to the declared combined standard uncertainty of 4.3~\muG. However, an \emph{a priori} approach to implement a SAC of ($+0.6\pm0.1$)~\muG\ has been described.

Such results can be related to the measurements performed during all the previous and future comparisons, especially the International Comparisons of Absolute Gravimeters (ICAGs). The most recent published results of an ICAG (i.e. 2005~\cite{icag}) have been chosen as a reference in the following. 

In the case of the MPG-2 (not present at the ICAG~2005) the computed SAC exceeds the preliminary declared uncertainty of 0.5~\muG\ due to this effect. Therefore, the results reported at previous comparisons by such AG can be updated with the obtained SAC, removing the contribution of 0.5~\muG\ from the uncertainty budget. 

About the IMGC-02, the application of the SAC of 0.6~\muG\ should not significantly change the \gi~values with respect to the combined standard uncertainty of about 4.3~\muG, which already included a contribution of 0.3~\muG\ for SAE~\cite{icag}. 

The results found for the FG5-238 can be considered for the other AGs of the same type (i.e. FG5-2$xx$) which represent the majority of the instruments participating to the comparisons. Usually the \gi\ measurements performed using FG5-$xxx$ did not present a SAC and only few participants considered the SAE in the uncertainty budget evaluation. At the ICAG~2005~\cite{icag}, a contribution of 0.1~\muG\ has been estimated for the instruments and it is has been used to calculate the so called \emph{conventional} uncertainty for all the FG5-$xxx$~\cite{vitu}. Hence, a revision of the values of the previous comparison is proposed, also because the abundance of FG5 instruments strongly influences the Key Comparison Reference Value.  

The SAE, computed using the FEM simulations, was treated as the additive perturbation of the Earth gravity field. The SAC was obtained as the mean-weighted value of the SAE over the measurement range. In the case of the linear SAE, the SAC corresponds to the SAE value, linearly interpolated to the effective measurement height (with the opposite sign). If the SAE significantly deviated from linearity, the SAC could be computed using the weighting function~\cite{nago1} of a particular instrument. The presented approach to evaluate the SAE and SAC is applied to three different AGs, but it has general validity. For this reason, the whole procedure can be applied to other absolute gravimeters, knowing its peculiar features as geometry, mass values and weighting functions. 

\ack
Special thanks to Filippo Greco and Antonio Pistorio of the Istituto Nazionale Geologia e Vulcanologia (INGV), Sezione di Catania, for the important collaboration in the characterisation of the FG5-238. Hua Hu from Tsinghua University, Beijing, People's Republic of China, has contributed to the characterization of the MPG-2. Contribution of the second author (SS) was supported by the Deutsche Forschungsgemeinschaft (DFG, Germany) under the project SV~86/1-1. Comments and suggestions of the referees were helpful to improve the manuscript.

\end{document}